\documentclass[twocolumn,preprintnumbers,aps,showpacs,superscriptaddress, pra]{revtex4-1}
\usepackage{palatino}
\usepackage{natbib}
\usepackage[latin1]{inputenc}
\usepackage{epsf}
\usepackage{amsmath,amssymb}
\usepackage{latexsym}
\usepackage{calc}
\usepackage{color}
\usepackage{epsfig}

\usepackage{hyperref}

\newcommand{\ben}{\begin{equation}}
\newcommand{\een}{\end{equation}}
\newcommand{\bea}{\begin{eqnarray}}
\newcommand{\eea}{\end{eqnarray}}

\def\bA{{\bf A}}

\def\dulR{{\underline{\underline{\bf R}}}}
\def\dulr{{\underline{\underline{\bf r}}}}

\begin{document}
\title{Laser-induced electron localization in  H$_2^+$: Mixed quantum-classical dynamics based on the exact time-dependent potential energy surface } 
\author{Yasumitsu Suzuki}
\altaffiliation[Present address: ]{Department of Physics, Tokyo University of Science, 1-3 Kagurazaka, Shinjuku-ku, Tokyo 162-8601, Japan}
\affiliation{Max-Planck Institut f\"ur Mikrostrukturphysik, Weinberg 2, D-06120 Halle, Germany}  
\author{Ali Abedi}
\altaffiliation[Present address: ]{Nano-Bio Spectroscopy Group, Departamento Fisica de Materiales, Universidad del Pais Vasco,
Centro de Fisica de Materiales CSIC-UPV/EHU-MPC and DIPC, Avenida Tolosa 72, E-20018 San Sebastian, Spain}
\affiliation{Max-Planck Institut f\"ur Mikrostrukturphysik, Weinberg 2, D-06120 Halle, Germany}
\affiliation{Department of Physics and Astronomy, Hunter College and the City University of New York, 695 Park Avenue, New York, New York 10065, USA}
\author{Neepa T. Maitra} 
\affiliation{Department of Physics and Astronomy, Hunter College and the City University of New York, 695 Park Avenue, New York, New York 10065, USA}
\author{E.K.U. Gross}
\affiliation{Max-Planck Institut f\"ur Mikrostrukturphysik, Weinberg 2, D-06120 Halle, Germany}
\affiliation{European Theoretical Spectroscopy Facility (ETSF)}

\date{\today}
  
\begin{abstract}
We study the exact nuclear time-dependent potential energy surface
(TDPES) for 
laser-induced electron localization with a view to eventually developing a mixed quantum-classical dynamics method for strong-field processes.  
The TDPES is defined
within the framework of the exact factorization [A. Abedi,
  N. T. Maitra, and E. K. U. Gross, Phys. Rev. Lett. 105, 123002
  (2010)] and contains the exact effect of the couplings to the electronic subsystem and to
any external fields within a scalar potential.  We compare its
features with those of the quasistatic potential energy surfaces
(QSPES) often used to analyse strong-field processes. We show that the
gauge-independent component of the TDPES has a mean-field-like
character very close to the density-weighted average of the
QSPESs. Oscillations in this component are smoothened out by the 
gauge-dependent component, and both components are needed to yield the
correct force on the nuclei.   Once the localization begins to set in,
the gradient of the exact TDPES tracks one QSPES and then switches to
the other, similar to the description provided by surface-hopping
between QSPESs. 
We show that evolving an ensemble of classical nuclear trajectories on the exact TDPES accurately 
reproduces the exact dynamics. 
This study suggests that the mixed quantum-classical dynamics scheme
based on evolving multiple classical nuclear trajectories on the exact 
TDPES will be a novel  and useful method to simulate strong field processes.
\end{abstract}

\maketitle 

\section{Introduction}
With the advent of attosecond technology~\cite{atto0,atto1,atto2,atto3,atto4,atto5}, the experimentally accessible time-scale has shifted  to that of electronic motion.
It allows the observation of electronic motion in real-time,
and even offers the control of electron motion and localization via lasers.
Several groups~\cite{sansone,HRB,Ray, Singh, Fischer, Calvert, KSIV, He2012, Liu, Jia, Zhuo,
Kling, Kremer, Roudnev, Tong, Graefe,  Rathje, Nora, Hli, Midorikawa} have demonstrated that it is
possible to control electronic motion in a dissociating molecule and
localize it selectively on one of the products of dissociation, with
several different strategies.  One technique employs 
the carrier envelope phase (CEP) 
of a single few-cycle laser pulse
~\cite{Kling, Kremer, Roudnev, Tong, Graefe, Rathje, Nora, Hli} , and another employs the time-delay between two
coherent ultrashort pulses~\cite{sansone, HRB, Ray, Singh, Fischer, Calvert, KSIV, He2012, Liu, Jia, Zhuo}.

These experiments so far treat small systems (such as H$_2$ and
D$_2$), with the aim of understanding the mechanisms of localization,
before applying the techniques to the control of larger systems~\cite{Kubel, Hli2}. 
Theoretical studies have a dual role~\cite{sansone, HRB, Ray, Singh, Fischer, Calvert, KSIV, 
He2012, Liu, Jia, Zhuo, Kling, Kremer, Roudnev, Tong, Graefe, Rathje, Nora, Hli, Midorikawa}
: (i) to help understand the complex correlation between the electron
dynamics and nuclear dynamics, and (ii) to establish  methods, generally
extendable to larger systems, that accurately simulate
the coupled electron-nuclear dynamics.  
For systems with more than two or three degrees of freedom, we must rely on
approximate methods, and usually some kind of mixed quantum-classical
approach is appropriate, where the electrons are treated
quantum-mechanically, coupled to nuclei described via classical
trajectories~\cite{Mitric, Ivano, Gonzalez, Gonzalez2}.  Different mixed quantum-classical schemes such as
Ehrenfest and surface-hopping~\cite{tully1,tully2,tully3}, differ in their treatment of the
classical nuclear motion, but use the same form for the potential
acting on the electrons.  For dynamics in strong fields, a
surface-hopping scheme between quasi-static potential energy surfaces
(QSPES) was introduced~\cite{Dietrich2,TIW,TIW2}, and in fact
applied to the electron-localization problem~\cite{KSIV}. Although
this semiclassical approach was shown to reproduce the experimental
asymmetries reasonably well, it is not altogether clear why
surface-hopping should give good predictions, given its problems associated with
over-coherence~\cite{Subotnik, Truhlar, Bittner, Prezhdo, Curchod, Makri, Kapral, Persico}.

In this paper, we will study the possibility of using a potential
derived from first-principles, the time-dependent potential energy
surface (TDPES)~\cite{AMG,AMG2}, in a mixed quantum-classical
description of the coupled dynamics. This potential arises out of the
exact factorization framework where a time-dependent Schr\"odinger
equation (TDSE) for the nuclei alone can be formulated. The potentials
appearing in this equation capture exactly all coupling to the
electronic system as well as any external fields, and the resulting
nuclear wavefunction reproduces the exact nuclear dynamics.  The
scalar potential is denoted the TDPES, and in many situations,
including all one-dimension problems, the TDPES is the {\it only}
potential acting on the nuclear subsystem; its gradient therefore
yields the exact force on the nuclei. For this reason, it is important
to gain an understanding of its structure, to address both points (i)
and (ii) above.  Therefore, our aim in this paper is to find the exact
TDPES for the problem of laser-induced electron localization in a one-dimensional model of H$_2^+$, 
compare its structure with potential surfaces more traditionally used for
strong-field dynamics, and study classical nuclear dynamics on the exact TDPES with a view to developing mixed quantum-classical schemes based on the exact factorization. 

Previous work~\cite{AAYG, AAYG2, AAYMMG} has analysed the structure
of the exact TDPES for a case of field-free dynamics, non-adiabatic
charge-transfer in the Shin-Metiu model~\cite{Metiu}, finding that
much intuition is gained by analysing it in term of the
Born-Oppenheimer (BO) potential energy surfaces (BOPESs), and that such
an analysis enables connections to be made with traditional
approximate methods for coupled electron-ion dynamics, such as surface-hopping.  Further, it was found that
evolving an ensemble of classical nuclear trajectories on the exact
TDPES accurately reproduces the exact nuclear
dynamics~\cite{AAYMMG}.

We will show here that analogous conclusions can be drawn for the
laser-induced electron localization problem: an ensemble of classical
nuclear trajectories evolving on the exact TDPES accurately reproduces
the exact nuclear dynamics, and analysis in terms of the QSPESs, which
play the role of the BOPESs when strong fields are present, is
helpful.  The TDPES naturally separates into a gauge-independent
part and a gauge-dependent part.  We show that the density-weighted
average of the QSPESs approximates the gauge-independent component,
which is rather oscillatory and the force on the nuclei resulting from
its gradient is incorrect.  Once the gauge-dependent component of the
TDPES is included, the oscillations smoothen out: together, they yield
the correct force on the nuclei.  Further, we find that, once
localization begins to set in, the gradient of the exact TDPES at the
location of the mean nuclear position, tracks that of one QSPES and then
switches to the other, resembling the picture provided by the
semiclassical surface-hopping approach~\cite{KSIV, TIW, TIW2}. 

A multiple trajectory Ehrenfest dynamics simulation shows that
although the nuclear dynamics is reasonably reproduced, an incorrect
electron localization asymmetry is obtained.  The error can be related
to the incorrect BO projections of the electronic wavefunction.  The
fact that the Ehrenfest dynamics yields inaccurate electron dynamics
can be anticipated from our recent work on the exact {\it
  electronic-}TDPES~\cite{SAMYG}: in this complementary picture,
instead of asking what is the exact potential acting on nuclei in an
exact TDSE for nuclei, one asks what is the exact potential acting on
electrons in an exact TDSE for the electronic subsystem. We
found~\cite{SAMYG} that the exact electronic-TDPES is significantly
different from the potential acting on electrons in the usual mixed
quantum-classical schemes -- including Ehrenfest as well as
surface-hopping schemes -- yielding significant errors in the
prediction of the electron localization asymmetry.  The results of the
present paper suggest that, instead, mixed quantum-classical schemes
based on evolving multiple classical trajectories on the exact TDPES
(or good approximations to it) will be a useful method to simulate
strong field processes.

This paper is organized as follows.  In section II, we review two
different concepts of potential energy surfaces for TD processes in
laser fields: the QSPES and the exact TDPES.  In section III we
compare the features of these potentials for electron localization
dynamics in the dissociation of a model H$_2^+$ molecule induced by
time-delayed coherent ultra shortlaser pulses.  We show the exact
TDPES gives the correct force acting on nuclei, so evolving multiple
classical trajectories on it reproduces the correct nuclear wavepacket
dynamics. The force obtained from surface-hopping between QSPESs can
approximately reproduce such an exact force once localization begins to set in.  We
also compute multiple trajectory Ehrenfest dynamics and reveal how it
fails to reproduce electron localization dynamics while it reasonably
reproduces the nuclear dynamics.  In section IV we summarize the
results and remark on the future directions.

\section{THEORY}

\subsection{Quasi-static potential energy surface}
\label{sec:QSPES}
In this section we first review the concept of the QSPES introduced
for the description of molecules in strong-fields. The QSPES has been
thoroughly discussed in earlier works~\cite{KSIV, Dietrich, Zuo, Seideman, 
Dietrich2,TIW,TIW2,Kono1998,Kono1999,Kono2004, Kono2009} 
, but we here give a discussion particularly relevant for the electron
localization dynamics problem in the dissociation of H$_2^+$.

For this problem, the essential physics is contained in
the two lowest field-free electronic states of the BO Hamiltonian, i.e., the 
1s$\sigma_g$ and 2p$\sigma_u$ states, and
 the full molecular wavefunction $\Psi(R,{\bf r},t)$ of the system can be expressed as
\ben
\Psi(R,{\bf r},t)=\chi_g(R,t)\phi^g_R({\bf r})
                +\chi_u(R,t)\phi^u_R({\bf r}) \;.
\label{eq:Psi-boexp}
\een
Here $\chi_g(R,t)$ and $\chi_u(R,t)$ describe nuclear
wavefunctions that exist in the 1s$\sigma_g$ and 2p$\sigma_u$ states respectively,
functions of the internuclear distance $R$ and time $t$,
and $\phi^g_R({\bf r})$ and $\phi^u_R({\bf r})$ describe 
the 1s$\sigma_g$ and 2p$\sigma_u$ electronic wavefunction respectively,
which parametrically depend on $R$.
Since $\phi^g_R({\bf r})$ and $\phi^u_R({\bf r})$ are 
bonding and anti-bonding combination of 1s atomic orbitals,
a coherent superposition of them provides the localized 
electronic states $\phi^{\rm left,right}_R({\bf r})=\frac{1}{\sqrt{2}}(\phi^g_R({\bf r})\pm \phi^u_R({\bf r}))$
that have the electron on either the left or the right proton.
These states form a convenient basis in which to monitor the electron localization asymmetry. 
In the experiment, interactions of the molecule with the time-delayed infra-red laser field in the
course of the dissociation provides a coupling of $\phi^g_R({\bf r})$ and $\phi^u_R({\bf r})$, creating a coherent superposition state, and, instead of Eq.~\ref{eq:Psi-boexp}, it is instructive to write:
\ben
\Psi(R,{\bf r},t)=\chi_{\rm left}(R,t)\phi^{\rm left}_R({\bf r})
                +\chi_{\rm right}(R,t)\phi^{\rm right}_R({\bf r}) 
\een
where $\chi_{\rm left}(R,t)$ and $\chi_{\rm right}(R,t)$  are defined as
the nuclear wavefunctions that exist in connection with
$\phi^{\rm left}_R({\bf r})$ and $\phi^{\rm right}_R({\bf r})$.
Measurements of ion fragment asymmetries left or right along the polarization
axis directly relate to $\chi_{\rm left}(R,t)$ and $\chi_{\rm right}(R,t)$.

While the field-free states above are useful to analyse the asymmetry,
to understand the time-development of the localization it is helpful to consider a third, 
time-dependent, basis,  the TD quasistatic states, $\phi^{\rm QS(i)}_R({\bf r},t)$, also known as phase-adiabatic states.
These states  are defined as instantaneous eigenstates of the  instantaneous electronic Hamiltonian 
 $\hat{H}^{\rm int}_R({\bf r},t)$, defined by
\ben
\hat{H}^{\rm int}_R({\bf r},t)=\hat{H}^{\rm BO}_R({\bf r})+\hat{v}_{\rm laser}({\bf r},t),
\een
i.e.,
\ben
\hat{H}^{\rm int}_R({\bf r},t)\phi^{\rm QS(i)}_R({\bf r},t)=\epsilon^{\rm QS(i)}(R,t)\phi^{\rm QS(i)}_R({\bf r},t)
\label{eq:QSPESdefn}
\een
where $\epsilon^{\rm QS(i)}(R,t)$ are the quasistatic potential energy surfaces (QSPESs).
Within our two-state model
we may write
\ben
\phi^{\rm QS(i)}_R({\bf r},t)=c_g^{(i)}(R,t)\phi^g_R({\bf r}) + c_u^{(i)}(R,t)\phi^u_R({\bf r}),
\een
so that the $\epsilon^{\rm QS(i)}(R,t)$ of Eq.~(\ref{eq:QSPESdefn})  are given by the eigenvalue equation:
\ben
\left(
    \begin{array}{cc}
      \langle \phi^g_R\arrowvert\hat{H}^{\rm int}_R\arrowvert\phi^g_R\rangle & \langle \phi^g_R\arrowvert\hat{H}^{\rm int}_R\arrowvert\phi^u_R\rangle  \\
      \langle \phi^u_R\arrowvert\hat{H}^{\rm int}_R\arrowvert\phi^g_R\rangle & \langle \phi^u_R\arrowvert\hat{H}^{\rm int}_R\arrowvert\phi^u_R\rangle
    \end{array}
\right)
\left(
    \begin{array}{cc}
      c_g^{(i)}   \\
      c_u^{(i)}  
    \end{array}
\right)   
  =\epsilon^{\rm QS(i)}
\left(
    \begin{array}{cc}
      c_g^{(i)}   \\
      c_u^{(i)}  
    \end{array}
  \right).  
\een  
Therefore we can express the QSPESs in terms of the BOPESs $\epsilon^{\rm BO(i)}(R)$  as
\ben
\begin{split}
\epsilon^{\rm QS(1,2)}(R,t)=&\epsilon^{\rm BO(1,2)}(R)\cos^2\theta(R,t) + \epsilon^{\rm BO(2,1)}(R)\sin^2\theta(R,t) \\
&\pm \langle \phi^g_R\arrowvert\hat{v}_{\rm laser}\arrowvert\phi^u_R\rangle \sin2\theta(R,t)
\end{split}
\een
and the electronic quasi-static eigenstates in terms of the BO states, 
\ben
\begin{split}
&\phi^{\rm QS(1)}_R({\bf r},t)=\cos\theta(R,t)\phi^g_R ({\bf r})+ \sin\theta(R,t)\phi^u_R({\bf r})\\
&\phi^{\rm QS(2)}_R({\bf r},t)=\sin\theta(R,t)\phi^g_R ({\bf r})- \cos\theta(R,t)\phi^u_R({\bf r}),
\end{split}
\een
where the TD mixing parameter $\theta(R,t)$ is given by
\ben
\tan2\theta(R,t)=\frac{2\langle \phi^g_R\arrowvert\hat{v}_{\rm laser}\arrowvert\phi^u_R\rangle}{\epsilon^{\rm BO(1)}(R)-\epsilon^{\rm BO(2)}(R)}.
\een
The molecular wavefunction expressed in terms of quasi-static states is
\ben
\Psi(R,{\bf r},t)=\chi^{\rm QS}_1(R,t)\phi^{\rm QS(1)}_R({\bf r},t)
                +\chi^{\rm QS}_2(R,t)\phi^{\rm QS(2)}_R({\bf r},t).
\label{eqn:exp_qs}                
\een
Note that the nuclear wavefunctions $\chi^{\rm QS}_1(R,t)$ and $\chi^{\rm QS}_2(R,t)$
that are connected to the quasi-static states $\phi^{\rm QS(1)}_R({\bf r},t)$ and $\phi^{\rm QS(2)}_R({\bf r},t)$
can be expressed in terms of $\chi_{\rm left}(R,t)$ and $\chi_{\rm right}(R,t)$ as
\ben
\begin{split}
\chi^{\rm QS}_1(R,t)=&\frac{1}{\sqrt{2}}[\chi_{\rm left}(R,t)(\cos\theta+\sin\theta)\\
&+\chi_{\rm right}(R,t)(\cos\theta-\sin\theta)] \\
\chi^{\rm QS}_2(R,t)=&\frac{1}{\sqrt{2}}[\chi_{\rm left}(R,t)(-\cos\theta+\sin\theta)\\
&+\chi_{\rm right}(R,t)(\cos\theta+\sin\theta)]. 
\end{split}
\een
which can be used to extract the electron localization from
$\chi^{\rm QS}_1(R,t)$ and $\chi^{\rm QS}_2(R,t)$.

A semi-classical surface-hopping model based on QSPESs
has recently been utilized to 
understand and reproduce the electron localization dynamics and asymmetry~\cite{KSIV,TIW,TIW2} in H$_2^+$. 
In this approach, an ensemble of classical nuclear trajectories evolve on one QSPES or the other QSPES,
making instantaneous hops between them as determined by a Landau-Zener formula. 
It was shown that the electron localization sets in a region where the dynamics is 
intermediate between adiabatic and diabatic: the ensemble of nuclear trajectories traverses 
several laser-induced avoided crossings between the QSPESs.
This semi-classical method gives asymmetry parameters in reasonably good overall agreement with that obtained from the full TDSE although the details differ.
The agreement lends some hope to the use of this semiclassical scheme to simulate coupled electron-ion
dynamics in control problems in more complicated systems; however, at the same time a
further understanding of the errors in the details is desirable. 
We will analyse this approach  by comparing the QSPESs with the exact TDPES, which we will review in the 
next section.

\subsection{Exact time-dependent potential energy surface}
In Ref.~\cite{AMG,AMG2}, it was shown that the 
full molecular wavefunction $\Psi(\dulr, \dulR,t)$ which
solves the TDSE
\ben
\hat{H}\Psi(\dulr, \dulR,t)=i\partial_t\Psi(\dulr, \dulR,t)
\een
can be exactly factorized to the single product
\ben
\Psi(\dulr, \dulR,t)=\chi(\dulR,t)\Phi_\dulR(\dulr,t)
\label{eqn: factorization}
\een
of the nuclear wavefunction $\chi(\dulR,t)$ and the
electronic wavefunction $\Phi_\dulR(\dulr,t)$ that 
parametrically depends on the
nuclear positions $\dulR$ and satisfies the partial normalization condition 
\ben
\int d\dulr |\Phi_\dulR(\dulr,t)|^2=1  \;\;\;\; \forall\dulR,t.
\label{eq:pnc}
\een
Here, the complete molecular Hamiltonian is 
\begin{equation}
 \hat{H} = \hat{T}_{n}(\dulR)+ \hat{V}^{n}_{ext}(\dulR,t) +\hat{H}_{\rm BO}(\dulr,\dulR)
 +\hat{v}^{e}_{ext}(\dulr,t),
\label{eq:completeH}
\end{equation} 
 and  $\hat{H}_{\rm BO}(\dulr,\dulR)$ is the  BO electronic Hamiltonian,
\begin{equation}
 \hat{H}_{\rm BO} = \hat{T}_{e}(\dulr)+ \hat{W}_{ee}(\dulr) +\hat{W}_{en}(\dulr,\dulR)+\hat{W}_{nn}(\dulR).
\label{eq:HBO}
\end{equation} 
Note that
$\hat{T}_{n}=-\sum_{\alpha=1}^{N_n}\frac{\nabla^2_\alpha}{2M_\alpha}$
and $\hat{T}_{e}=-\sum_{j=1}^{N_e}\frac{\nabla^2_j}{2m}$ are the
nuclear and electronic kinetic energy operators, $\hat{W}_{ee}$,
$\hat{W}_{en}$ and $\hat{W}_{nn}$ are the electron-electron, electron-nuclear and
nuclear-nuclear interaction, and $\hat{V}^{n}_{ext}(\dulR,t)$ and
$\hat{v}^{e}_{ext}(\dulr,t)$ are time-dependent (TD) external
potentials acting on the nuclei and electrons, respectively.
Throughout this paper $\dulR$ and $\dulr$ collectively represent the
nuclear and electronic coordinates respectively and $\hbar=1$. 

Returning to Eq.~(\ref{eqn: factorization}), the stationary variations of the quantum mechanical action with respect to 
$\Phi_\dulR(\dulr,t)$ and $\chi(\dulR,t)$ under the condition~(\ref{eq:pnc}) lead
 to the 
following equations of motion for $\chi(\dulR,t)$ and $\Phi_\dulR(\dulr,t)$:
\ben
\begin{split}
 \left(\hat{H}_{BO}(\dulr,\dulR)+\hat{v}^{e}_{ext}(\dulr,t)+\hat U_{en}^{coup}[\Phi_\dulR,\chi]-\epsilon(\dulR,t)\right)
 \Phi_{\dulR}(\dulr,t)\\
 =i\partial_t \Phi_{\dulR}(\dulr,t) 
\end{split}\label{eqn: exact electronic eqn}
 \een
\ben
\begin{split}
 \left[\sum_{\alpha=1}^{N_n} \frac{\left[-i\nabla_\alpha+\bA_\alpha(\dulR,t)\right]^2}{2M_\alpha} +\hat{V}^{n}_{ext}(\dulR,t) +
 \epsilon(\dulR,t)\right]\chi(\dulR,t)\\
 =i\partial_t \chi(\dulR,t) \label{eqn: exact nuclear eqn}.
\end{split}
 \een
Here, $\epsilon(\dulR,t)$ is the exact nuclear TDPES
\begin{equation}\label{eqn: tdpes}
 \epsilon(\dulR,t)=\left\langle\Phi_\dulR(t)\right|\hat{H}_{BO}+\hat{v}^{e}_{ext}(\dulr,t)+\hat U_{en}^{coup}-i\partial_t\left|
 \Phi_\dulR(t)\right\rangle_\dulr,
\end{equation}
$\hat U_{en}^{coup}[\Phi_\dulR,\chi]$ is the ``electron-nuclear coupling operator'',
\begin{align}
\hat U_{en}^{coup}&[\Phi_\dulR,\chi]=\sum_{\alpha=1}^{N_n}\frac{1}{M_\alpha}\left[
 \frac{\left[-i\nabla_\alpha-\bA_\alpha(\dulR,t)\right]^2}{2} \right.\label{eqn: enco} \\
& \left.+\left(\frac{-i\nabla_\alpha\chi}{\chi}+\bA_\alpha(\dulR,t)\right)
 \left(-i\nabla_\alpha-\bA_{\alpha}(\dulR,t)\right)\right],\nonumber
\end{align}
and $\bA_{\alpha}\left(\dulR,t\right)$ is the TD vector potential potential,
\begin{equation}\label{eqn: vector potential}
 \bA_{\alpha}\left(\dulR,t\right) = \left\langle\Phi_\dulR(t)\right|-i\nabla_\alpha\left.\Phi_\dulR(t)
 \right\rangle_\dulr.
\end{equation}
The symbol $\left\langle\,\,\cdot\,\,\right\rangle_\dulr$ indicates an 
integration over electronic coordinates only. 
Note that the PNC makes the factorization~(\ref{eqn: factorization}) unique up to within a $(\dulR,t)$-dependent gauge transformation, 
\begin{equation}\label{eqn: gauge}
 \begin{array}{rcl}
  \chi(\dulR,t)\rightarrow\tilde\chi(\dulR,t)&=&e^{-i\theta(\dulR,t)}\chi(\dulR,t) \\
  \Phi_\dulR(\dulr,t)\rightarrow\tilde\Phi_\dulR(\dulr,t)&=&e^{i\theta(\dulR,t)}\Phi_\dulR(\dulr,t),
 \end{array}
\end{equation}
and Eqs. (\ref{eqn: exact electronic eqn}) and (\ref{eqn: exact nuclear eqn}) are form invariant under this transformation while 
the scalar potential and the vector potential transform as  
\begin{eqnarray}
\tilde{\epsilon}(\dulR,t) &=& \epsilon(\dulR,t)+\partial_t\theta(\dulR,t)\label{eqn: transformation of epsilon} \\
\tilde{\bf A}_{\alpha}(\dulR,t) &=& {\bf A}_{\alpha}(\dulR,t)+\nabla_\alpha\theta(\dulR,t)\label{eqn: transformation of A}.
\end{eqnarray}

The equation for the exact nuclear wavefunction, Eq.~(\ref{eqn: exact nuclear eqn}), is Schr\"odinger-like, and the  TD vector
potential~(\ref{eqn: vector potential}) and  TD scalar 
potential~(\ref{eqn: tdpes}) that appear in it, exactly govern the nuclear dynamics.
It is important to note that $\chi(\dulR,t)$ can be interpreted 
as the exact nuclear wave-function since it leads to an $N$-body nuclear density, 
$\Gamma(\dulR,t)=\vert\chi(\dulR,t)\vert^2,$ and an $N$-body current density, 
${\bf J}_\alpha(\dulR,t)=\frac{1}{M_\alpha}\Big[\mbox{Im}(\chi^*(\dulR,t)\nabla_\alpha\chi(\dulR,t))+
\Gamma(\dulR,t){\bf A}_\alpha(\dulR,t)\Big],$  which reproduce the true nuclear 
$N$-body density and current density~\cite{AMG2} obtained from the full
wave-function $\Psi(\dulr,\dulR,t)$. 

In our previous work the shape of this exact TDPES has been useful to
interpret dynamics for both a strong field process (strong-field
dissociation of H$_2^+$)~\cite{AMG,AMG2} as well as for field-free
dynamics of non-adiabatic charge-transfer~\cite{AAYG,AAYG2,AAYMMG}.
In particular, in the field-free case, a detailed study of the form of
its gauge-dependent and gauge-independent parts proved instructive to
understand its effect on the nuclear dynamics, and the structure to be
expected for general field-free problems.  Importantly, in a mixed
quantum-classical description, the gradient of this exact TDPES gives
uniquely the correct force on the nuclei, and it was shown, in the
field-free case, that an ensemble of classical trajectories evolving
on the exact TDPES accurately reproduces the exact nuclear wavepacket
dynamics.  We now consider a detailed study of the form of the exact
TDPES for the present case of dynamics in external fields, with the
aims of addressing three questions. First, does running
classical nuclear dynamics on the exact TDPES reproduce the dynamics
of laser-induced electron localization? Second, how are the QSPESs related
to the exact TDPES? Third, can we see hints of the semiclassical
surface-hopping method in the exact TDPES? 

\section{Results and discussion}

\subsection{Theoretical model}
We employ a one-dimensional model of the H$_2^+$ molecule to study
electron localization dynamics  achieved by time-delayed coherent ultra short
laser pulses~\cite{sansone,HRB,KSIV}. In the experiment,
first an ultraviolet (UV) pulse excites H$_2^+$ to the 
dissociative 2p$\sigma_u$ state while a second
time-delayed infrared (IR) pulse induces electron
transfer between the dissociating atoms.
In our model, we start
the dynamics after the excitation by the UV pulse:
the wavepacket starts at $t=0$ on the first excited state (2p$\sigma_u$ state)
of H$_2^+$ as a Frank-Condon projection of the wavefunction
of the ground state, and then is exposed to the IR laser pulse.
The full Hamiltonian of the system is given by
\ben
\begin{split}
&\hat{H}(R,z,t)=\hat{T}_n(R)+\hat{H}^{\rm int}_{R}(z,t)\\
&=\hat{T}_n(R)+\hat{T}_e(z)+\hat{W}_{nn}(R)+\hat{W}_{en}(z,R)+\hat{v}_{\rm laser}(z,t)
\end{split}
\een
where $R$ is the internuclear distance and $z$ is the electronic coordinate as
measured from the nuclear center of mass.
The kinetic energy terms are $\hat{T}_n(R) = -\frac{1}{2\mu_n}\frac{\partial^2}{\partial R^2}$ and, 
$\hat{T}_e(z) = -\frac{1}{2\mu_e}\frac{\partial^2}{\partial z^2}$, respectively, where the reduced mass of the nuclei is given by $\mu_n=M_{\rm H}/2$, 
and reduced electronic mass is given by $\mu_e=\frac{2M_{\rm H}}{2M_{\rm H}+1}$ ($M_{\rm H}$ is the proton mass).
The interactions are soft-Coulomb: $\hat{W}_{nn}(R) = \frac{1}{\sqrt{0.03+R^2}}$,
and $\hat{W}_{en}(z,R) = -\frac{1}{\sqrt{1.0+(z-\frac{R}{2})^2}} -\frac{1}{\sqrt{1.0+(z+\frac{R}{2})^2}}$ (and $\hat{W}_{ee} = 0$).
The IR pulse is described within the dipole approximation and length gauge, as $\hat{v}^e_{ext}(z,t) = E(t)q_ez$,
where $E(t)=E_0\exp\left[ -\left( \frac{t-\Delta t}{\tau}\right) ^2\right]\cos(\omega (t-\Delta t))$, and
 the reduced charge $q_e=\frac{2M_{\rm H}+2}{2M_{\rm H}+1}$. The wavelength is 800 nm and the peak intensity $I_0=E_0^2=3.0\times10^{12}$W/cm$^2$. 
The pulse duration  is $\tau =4.8 fs$ and $\Delta t$ is the time delay between the UV and IR pulses. 
Here we show the results of $\Delta t=$ 7 fs.

\begin{figure}[h]
 \centering
 \includegraphics*[width=1.0\columnwidth]{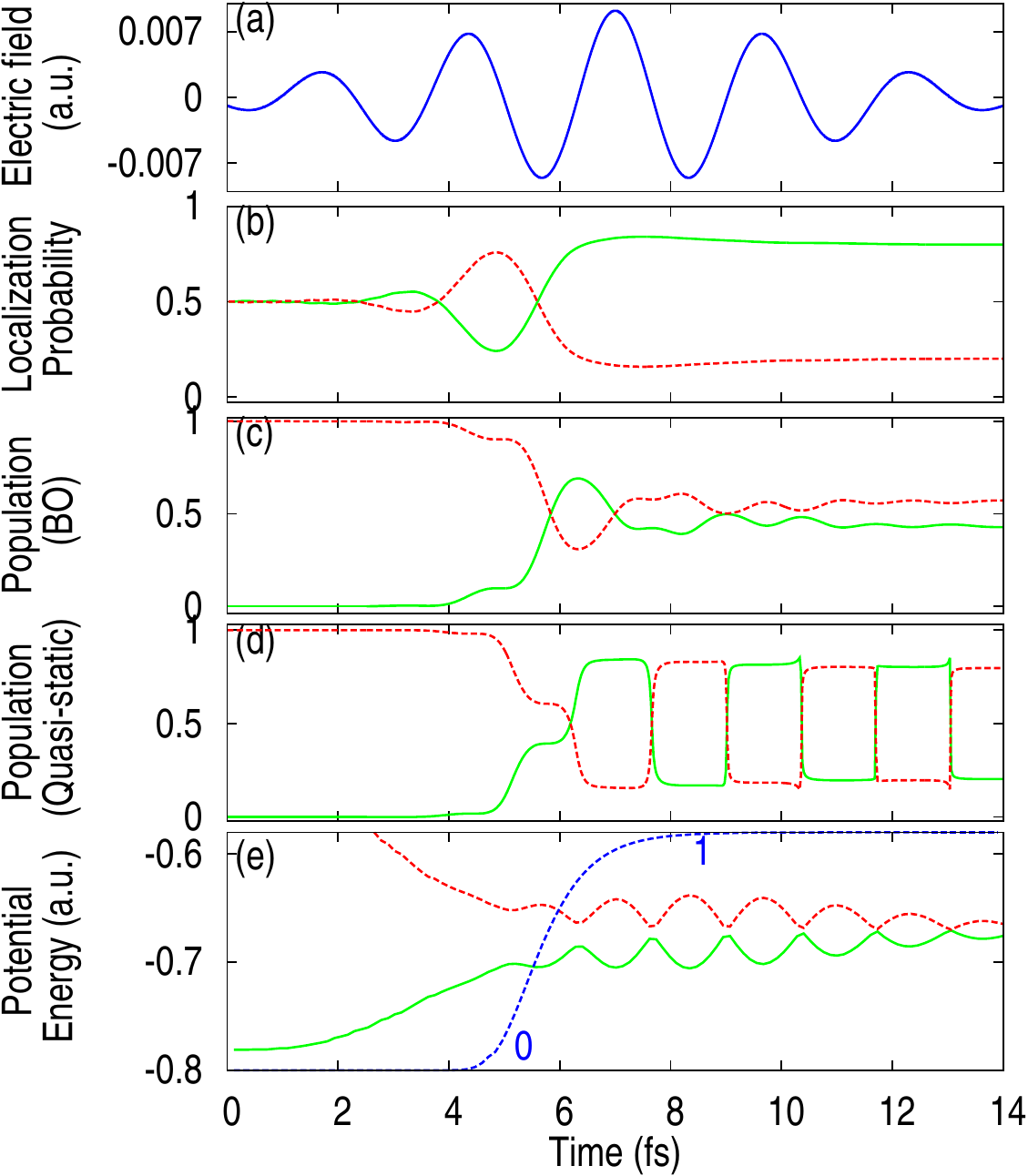}
 \caption{(a) 4.8 fs FWHM 800 nm laser pulse. (b) Electron localization probabilities along the negative (green solid line) and 
 the positive z-axis (red dashed line) as a function of time. (c) Population dynamics
 during dissociation on the BO state $\phi^g_R(z)$ (green solid) and $\phi^u_R(z)$ (red dashed).
 (d) Population dynamics
 during dissociation on the 1st quasi-static state  $\phi^{\rm QS(1)}_R(z,t)$ (green solid) 
 and 2nd quasi-static state $\phi^{\rm QS(2)}_R(z,t)$ (red dashed).
 (e) Quasi-static potential energy surfaces $\epsilon^{\rm QS(1)}(R,t)$ (green solid) and $\epsilon^{\rm QS(2)}(R,t)$ (red dashed)
 for a nuclear trajectory $\langle R \rangle(t)$ that tracks 
 the expectation value of the internuclear distance.
 The blue curve shows the transition probability given by a Landau-Zener
 formula~ (Eq. 18 of Ref.~\cite{KSIV}).}
 \label{fig:Fig1}
\end{figure}

We propagate the full TDSE
\ben
\hat{H}(z,R,t)\Psi(z,R,t)=i\partial_t\Psi(z,R,t)
\een
numerically exactly to obtain the full molecular wavefunction
$\Psi(z,R,t)$, and from it we calculate
the probabilities of directional localization of the electron, $P_{\pm}$, 
which are defined as $P_{+(-)} = \int_{z>(<)0} dz \int dR |\Psi(z,R,t)|^2$.
These are shown as the green solid ($P_-$) and red dashed ($P_+$) lines in Fig.~\ref{fig:Fig1}b.
It is evident from this figure that considerable electron localization occurs, with the electron density predominantly 
localized on the left (negative z-axis).

Furthermore, we calculate the population dynamics 
of the BO states $\phi^g_R(z)$ (green solid) and $\phi^u_R(z)$ (red
dashed) (Fig.~\ref{fig:Fig1}c) during dissociation, as well as the population dynamics on the 1st
quasi-static state $\phi^{\rm QS(1)}_R(z,t)$ (green solid) and 2nd
quasi-static state $\phi^{\rm QS(2)}_R(z,t)$ (red dashed)
(Fig.~\ref{fig:Fig1}d); the relative simplicity of the latter demonstrate the usefulness of the QS basis for laser-induced processes. 
  We then plot the QSPESs $\epsilon^{\rm
  QS(1)}(\langle R(t)\rangle,t)$ (green solid) and $\epsilon^{\rm QS(2)}(\langle R(t)\rangle,t)$ (red dashed)
evaluated at a nuclear trajectory $\langle R(t) \rangle$ that tracks
the expectation value of the internuclear distance.  These results
coincide qualitatively with the previous results reported by
Kelkensberg et al.~\cite{KSIV} 
Panels b, d, and e, suggest that the electron
localization is determined by the passage of the
dissociating molecule through a regime where the laser-molecule
interaction is neither diabatic nor adiabatic.  As discussed in the
previous section, the semiclassical scheme, with the avoided crossings between the QSPES 
inducing the trajectories to hop between them, reproduces the general behavior.  
Next, we will compare the exact TDPES with the QSPES to understand the relation
between the two, shed some light on the surface-hopping scheme, and find the exact force on classical nuclei.

\subsection{Exact TDPES vs. QSPES}
\begin{figure}[h]
 \centering
 \includegraphics*[width=1.0\columnwidth]{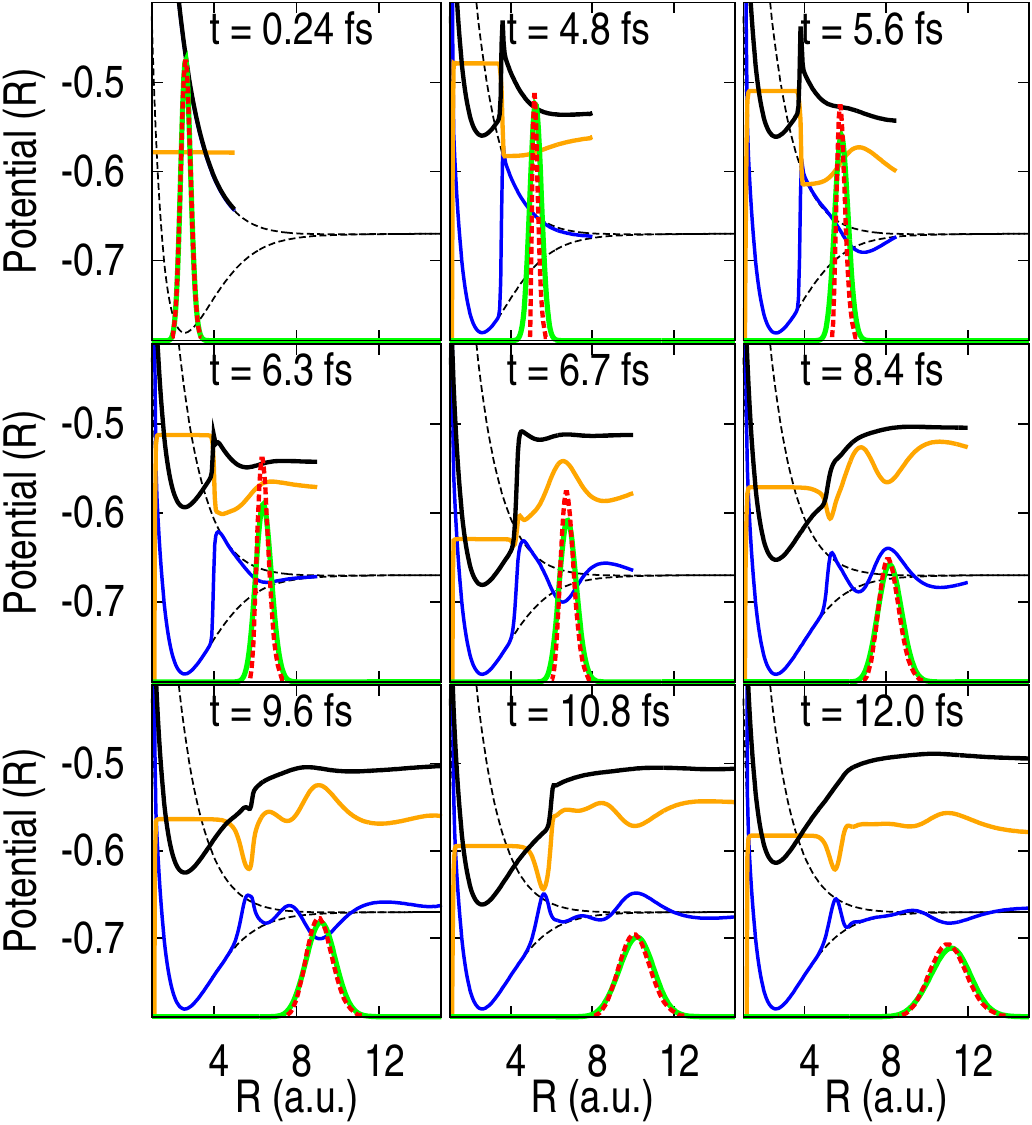}
 \caption{Snapshot of the exact TDPES $\epsilon(R,t)$ (black solid)
 , its gauge-invariant part $\epsilon_{gi}(R,t)$ (blue solid) and 
 gauge-dependent part $\epsilon_{gd}(R,t)$ (orange solid) at indicated times
 along with two lowest BOPESs (black dashed).
 Furthermore, the exact nuclear density $|\chi(R,t)|^2$ 
 (green solid) and the nuclear density reconstructed from
 the multiple trajectory dynamics on the exact TDPES (red dashed)
 for each time are also plotted.}
 \label{fig:Fig2}
\end{figure}

First we show the exact TDPES for this process in Fig.~\ref{fig:Fig2}.
We calculate the TDPES in the gauge where the vector
potential $A(R,t)$ is zero~\cite{AMG2}, so the TDPES $\epsilon(R,t)$ is the only
potential acting on the nuclear subsystem.
It is instructive to express the TDPES  as the sum of 
the gauge-independent term $\epsilon_{gi}(R,t)$ and the gauge-dependent term $\epsilon_{gd}(R,t)$
as done in previous studies~\cite{AMG2, AAYG} :
\ben
\epsilon(R,t)=\epsilon_{gi}(R,t)+\epsilon_{gd}(R,t)
\een
where 
\ben
\epsilon_{gi}(R,t)=\left\langle\Phi_R(t)\right|\hat{H}_{BO}+\hat{v}_{\rm laser}+\hat U_{en}^{coup}\left|
\Phi_R(t)\right\rangle_z
\een
and
\ben
\epsilon_{gd}(R,t)=\left\langle\Phi_R(t)\right|-i\partial_t\left|
\Phi_R(t)\right\rangle_z.
\een 
In Fig.~\ref{fig:Fig2}, $\epsilon(R,t)$(black solid), $\epsilon_{gi}(R,t)$(blue solid) and $\epsilon_{gd}(R,t)$(orange solid)
are plotted at nine different times, along with the two lowest BOPESs, $\epsilon^{\rm BO (1)}(R)$ and $\epsilon^{\rm BO (2)}(R)$. 
(Note that the TDPES $\epsilon(R,t)$(black solid) and its GD component $\epsilon_{gd}(R,t)$(orange solid)
have been rigidly shifted along the energy axis).

We also plot
the exact nuclear density $|\chi(R,t)|^2$ (green solid) and 
the nuclear density reconstructed from evolving an ensemble of $800$
 classical trajectories on the exact TDPES (red dashed)~\cite{AAYMMG}
 at each time. 
The closeness of these last two curves  shows that a mixed quantum-classical scheme 
for the electron localization process is appropriate and that the exact TDPES $\epsilon(R,t)$
gives the correct force acting on classical nuclei in such a scheme.

In previous work~\cite{AAYG,AAYG2,AAYMMG}, step-like features of
$\epsilon_{gi}(R,t)$ and $\epsilon_{gd}(R,t)$ in the field-free
non-adiabatic process in the vicinity of the avoided crossing have
been shown. In particular, after passage through the avoided crossing, where the nuclear 
wavepacket had spatially separated on two BOPESs, the GI component tracked one BO surface
or the other, with a step between them, while the GD component was piecewise flat, but 
with a step in the same region with opposite sign. The net TDPES was overall more smooth than either of the components. Here, 
we find again very interesting features of $\epsilon_{gi}(R,t)$ and $\epsilon_{gd}(R,t)$. First note that both 
$\epsilon_{gi}(R,t)$ and $\epsilon_{gd}(R,t)$ shows many small hills and valleys after the
laser-induced nonadiabatic transitions begin, but with opposite slopes to each other,
so that these structures largely cancel each other when the exact TDPES
$\epsilon(R,t)$ is constructed (much like the near-cancellation of the steps in the field-free case).
Like the field-free case, both the GI and GD terms are important to consider
to predict the correct nuclear dynamics. 
Second, in the present strong-field case, unlike the field-free examples studied in~\cite{AAYG,AAYG2,AAYMMG},  
$\epsilon_{gi}(R,t)$ does not piecewise track one BOPES or the other.
However, it does track a {\it density-weighted} QSPES, as we will show next. 

\begin{figure}[h]
 \centering
 \includegraphics*[width=1.0\columnwidth]{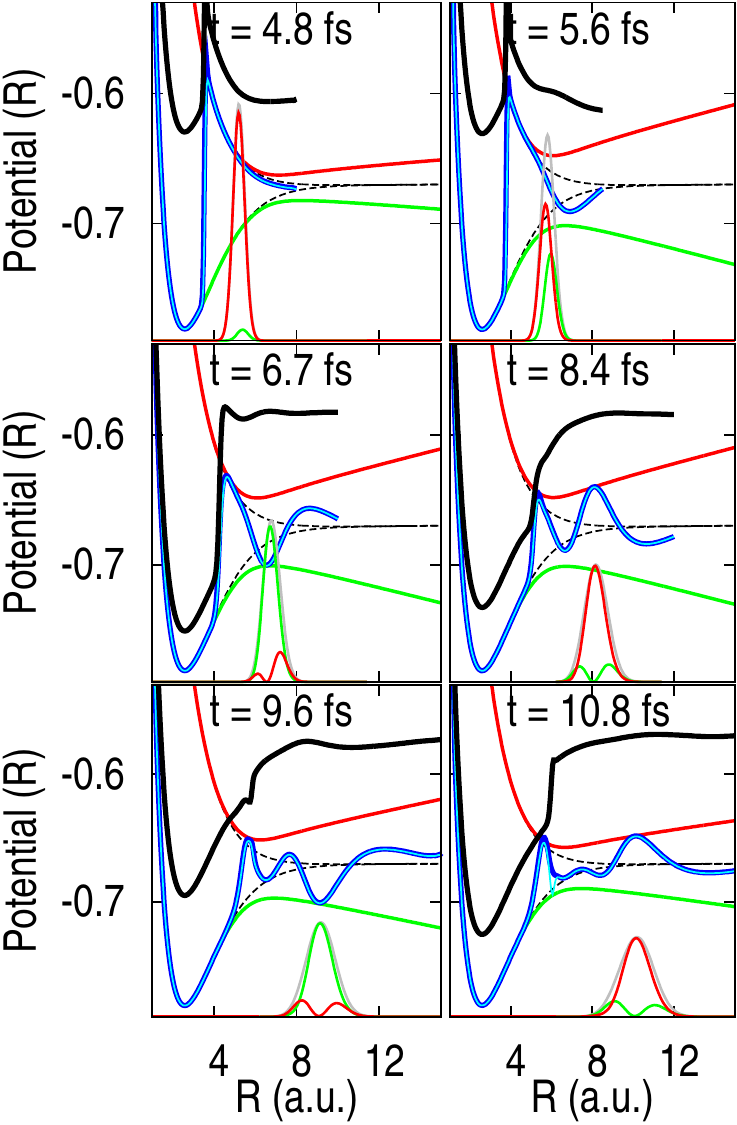}
 \caption{Snapshots of the gauge-invariant part of exact TDPES $\epsilon_{gi}(R,t)$ (blue solid), 
 QSPESs $\epsilon^{\rm QS(1)}(R,t)$ (green solid) and $\epsilon^{\rm QS(2)}(R,t)$ (red solid),
and the weighted average of the QSPESs $\epsilon^{\rm QS}_{ave}(R,t)$ (light blue solid) at indicated times. 
  $|\chi^{\rm QS}_1(R,t)|^2$ (green) and  $|\chi^{\rm QS}_2(R,t)|^2$ (red) are also plotted.}
 \label{fig:Fig3}
\end{figure}

In Fig.~\ref{fig:Fig3}, we show $\epsilon(R,t)$(black solid)
(which is again rigidly shifted along the energy axis) and the gauge-invariant part 
 $\epsilon_{gi}(R,t)$ (blue solid) together with the QSPESs 
$\epsilon^{\rm QS(1)}(R,t)$ (green solid) and $\epsilon^{\rm
  QS(2)}(R,t)$ (red solid).  
We find that the oscillations in the
gauge-invariant part of exact TDPES $\epsilon_{gi}(R,t)$ (blue solid)
tend to step between the two QSPESs: $|\chi^{\rm
  QS}_1(R,t)|^2$ and $|\chi^{\rm QS}_2(R,t)|^2$ are also plotted in
Fig.~\ref{fig:Fig3}, and we see that $\epsilon_{gi}(R,t)$ tends
towards the QSPES whose population is dominant, i.e. when $|\chi^{\rm
  QS}_1(R,t)|^2$ is larger than $|\chi^{\rm QS}_2(R,t)|^2$
$\epsilon_{gi}(R,t)$ approaches to $\epsilon^{\rm QS(1)}(R,t)$ and
when $|\chi^{\rm QS}_2(R,t)|^2$ is larger than $|\chi^{\rm
  QS}_1(R,t)|^2$ $\epsilon_{gi}(R,t)$ approaches to $\epsilon^{\rm
  QS(2)}(R,t)$. In fact, $\epsilon_{gi}(R,t)$ lies practically on top of the 
the weighted average of the quasi-static surfaces $\epsilon^{\rm QS}_{ave}(R,t)$:
\ben
\begin{split}
\epsilon^{\rm QS}_{ave}(R,t)&=\frac{|\chi^{\rm QS}_1(R,t)|^2}{|\chi^{\rm QS}_1(R,t)|^2+|\chi^{\rm QS}_2(R,t)|^2}\epsilon^{\rm QS(1)}(R,t)\\
&+\frac{|\chi^{\rm QS}_2(R,t)|^2}{|\chi^{\rm QS}_1(R,t)|^2+|\chi^{\rm QS}_2(R,t)|^2}\epsilon^{\rm QS(2)}(R,t)
\label{eq:epsQS_ave}
\end{split}
\een
This is plotted with light blue line in Fig.~\ref{fig:Fig3}.
Therefore the weighted-average of the QSPESs approximates 
the gauge-invariant part of exact TDPES $\epsilon_{gi}(R,t)$,
but not the full exact TDPES $\epsilon(R,t)$. 
In fact, this is quite analogous to the previous results on the field-free passage 
through an avoided crossing~\cite{AAYG,AAYG2,AAYMMG}: there, at the times considered,
the density-weighted average collapsed to one BO surface or the other except in
the intermediate (step) region, because the spatial separation of the parts of 
the density projected onto different BO surfaces meant that  in the field-free analog to Eq.~\ref{eq:epsQS_ave}, 
the prefactors of each term was either one or zero. Here it is evident that 
the density does not spatially separate (Fig.~\ref{fig:Fig3}), i.e. the projections on to 
the QSPESs overlap. One can make entirely analogous statements in both cases: 
the density-weighted average of the BOPES  approximates the gauge-invariant part 
of exact TDPES $\epsilon_{gi}(R,t)$ in the field-free case, and the density-weighted average of the QSPES  
approximates the gauge-invariant part of exact TDPES $\epsilon_{gi}(R,t)$ in the presence of strong fields.

To confirm the relationship between $\epsilon_{gi}(R,t)$ and $\epsilon^{\rm QS}_{ave}(R,t)$,
we consider the expansion of the complete wavefunction with the 
two lowest quasi-static states (Eq.~\ref{eqn:exp_qs}).
Then the exact electronic conditional wavefunction $\Phi_R(z,t)$ is expressed as:
\ben
\Phi_R(z,t)=\frac{\chi^{\rm QS}_1(R,t)}{\chi(R,t)}\phi^{\rm QS(1)}_R(z,t)
                +\frac{\chi^{\rm QS}_2(R,t)}{\chi(R,t)}\phi^{\rm QS(2)}_R(z,t).                
\een
Then we realize:
\ben
\begin{split}
&\langle\Phi_R(z,t)|\hat{H}^{\rm BO}+\hat{v}_{\rm laser}|\Phi_R(z,t)\rangle_z \\
&=\frac{|\chi^{\rm QS}_1(R,t)|^2}{|\chi(R,t)|^2}\epsilon^{\rm QS(1)}+\frac{|\chi^{\rm QS}_2(R,t)|^2}{|\chi(R,t)|^2}\epsilon^{\rm QS(2)}\\
&=\epsilon^{\rm QS}_{ave}(R,t).
\end{split}
\een
Since $\epsilon_{gi}(R,t)=\langle\Phi_R(z,t)|\hat{H}^{\rm BO}+\hat{v}_{\rm laser}|\Phi_R(z,t)\rangle_z
+\frac{1}{2M}\langle\Phi_R(z,t)\arrowvert (-i\frac{\partial}{\partial R}-A(R,t))^2  \arrowvert\Phi_R(z,t)\rangle_z$,
we can conclude
\ben
\epsilon_{gi}(R,t)\approx\epsilon^{\rm QS}_{ave}(R,t),
\een
because $O(M^{-1})$ term gives a much smaller contribution.

To reproduce the correct dynamics, however the effect of
$\epsilon_{gd}(R,t)$ is crucial to include, as in the field-free case
studied before~\cite{AAYMMG}. In the gauge we have chosen $A(R,t) = 0$, but we note that if instead we choose the
gauge where $\epsilon_{gd}(R,t)=0$ then the vector potential $A(R,t)$ will be non-zero, and 
will be responsible for the role of effectively reducing the
oscillatory structure in the GI term.

\begin{figure}[h]
 \centering
 \includegraphics*[width=1.0\columnwidth]{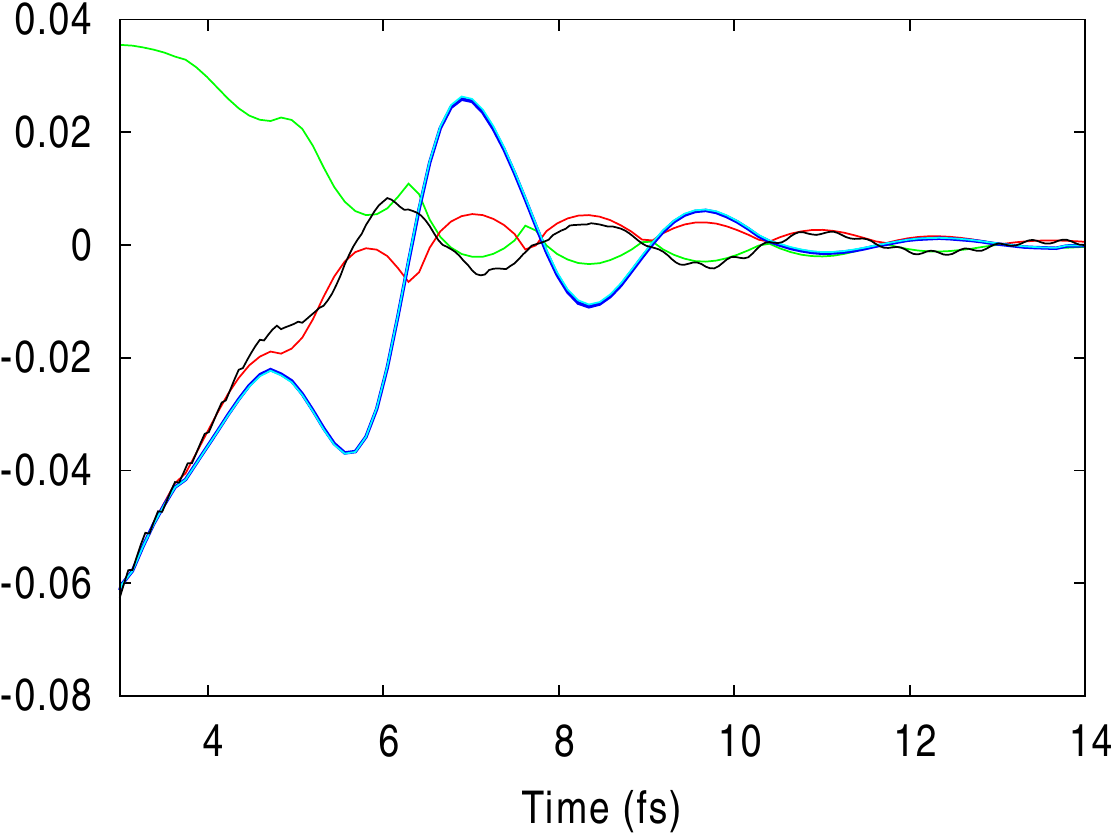}
 \caption{Time evolution of the gradient of each TDPES at position $\langle R(t)\rangle$.
 Green line: $\frac{\partial}{\partial R}\epsilon^{\rm QS(1)}(\langle R(t)\rangle)$;
 Red line: $\frac{\partial}{\partial R}\epsilon^{\rm QS(2)}(\langle R(t)\rangle)$;
 Blue line: $\frac{\partial}{\partial R}\epsilon_{gi}(\langle R(t)\rangle)$;
 Light blue line: $\frac{\partial}{\partial R}\epsilon^{\rm QS}_{ave}(\langle R(t)\rangle)$; 
 Black line: $\frac{\partial}{\partial R}\epsilon(\langle R(t)\rangle)$.}
 \label{fig:Fig4}
\end{figure}

In Fig.~\ref{fig:Fig4}, we plot the 
gradient of the different potentials computed on the trajectory of mean nuclear distance $\langle R\rangle(t)$, as a more direct probe of the force on the nuclei.
The black line, which is  the gradient of the exact TDPES
$\frac{\partial}{\partial R}\epsilon(\langle R(t)\rangle)$,  gives the exact force on the nuclei.
First we immediately notice that 
the gradient of the weighted average of the two QSPES
$\frac{\partial}{\partial R}\epsilon^{\rm QS}_{ave}(\langle
R(t)\rangle)$ (light blue line) (equivalently, the GI component (blue
line)) is completely different from the exact force. A semi-classical simulation on the
weighted average of the two QSPES would not give the correct nuclear
dynamics.  We observe instead that, as the localization sets in, the exact
force $\frac{\partial}{\partial R}\epsilon(\langle R(t)\rangle)$
coincides with the gradient of one or the other QSPES (red or green).
This supports the idea of semiclassical surface-hopping between
QSPES~\cite{TIW,TIW2,KSIV} at least after the localization begins to
set in (time $\sim$6 fs): the exact force on the nuclei is given by
the gradient of the exact TDPES, and, when evaluated at the mean
nuclear position, coincides with the force from one QSPES or the
other, making transitions between them at their avoided
crossings. This explains why the semiclassical simulations of
Ref.~\cite{KSIV} had a reasonable agreement with the exact results.
Furthermore the figure shows the important role of the gauge-dependent part $\epsilon_{gd}(R,t)$;  
without this term, the force on the nuclei would be more oscillatory and quite different (blue line in the figure). 
We note that if instead we choose the gauge where $\epsilon_{gd}(R,t)=0$ then the 
vector potential $A(R,t)$ will be responsible for the role of effectively reducing the oscillatory structure in the GI term.
As stated above, when we choose the gauge where $\epsilon_{gd}(R,t)=0$, then the 
vector potential $A(R,t)$ plays the role of it according to their relationship:
$\tilde A(R,t)=\int^t_0 dt'\left(-\partial_R\epsilon_{gd}(R,t')\right)$ \cite{AAYMMG}.

\subsection{Multiple trajectory Ehrenfest dynamics}

\begin{figure}[h]
 \centering
 \includegraphics*[width=1.0\columnwidth]{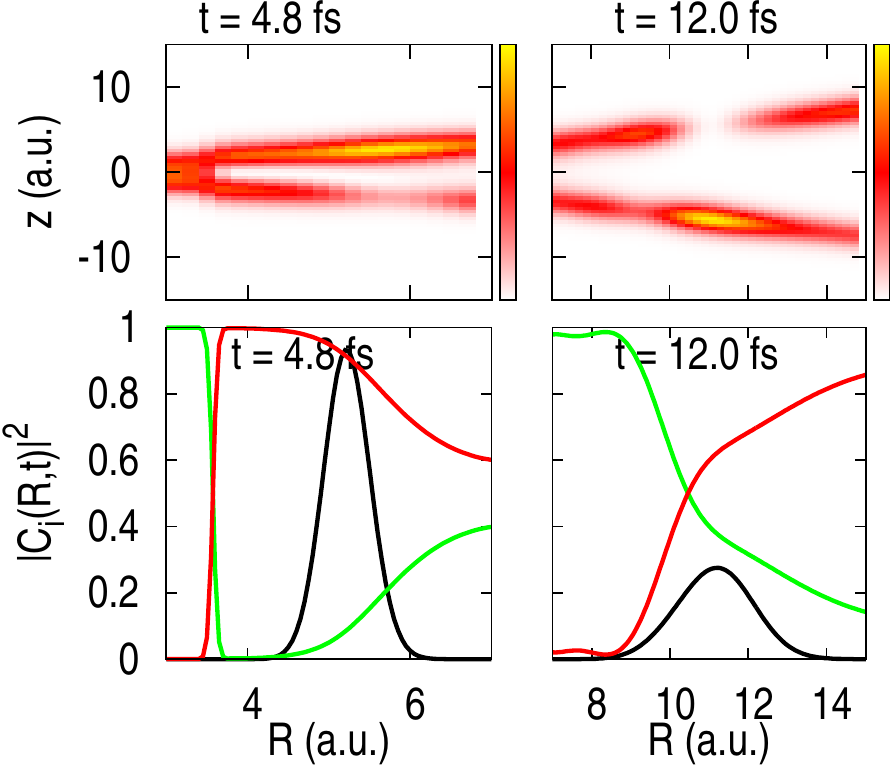}
 \caption{Upper panel: Conditional electron density $|\Phi_R(z,t)|^2$
 obtained from the exact calculation at the indicated times.
 Lower panel: Squared expansion coefficients of the Born-Oppenheimer expansion
 $|C_g(R,t)|^2$ (green) and $|C_u(R,t)|^2$ (red) of the exact conditional electronic wave function $\Phi_R(z,t)$
 ($\Phi_R(z,t)=C_g(R,t)\Phi^g_R(z)+C_u(R,t)\Phi^u_R(z)$) at the indicated times. The exact nuclear density is also plotted (black). }
 \label{fig:Fig5}
\end{figure}

\begin{figure}[h]
 \centering
 \includegraphics*[width=1.0\columnwidth]{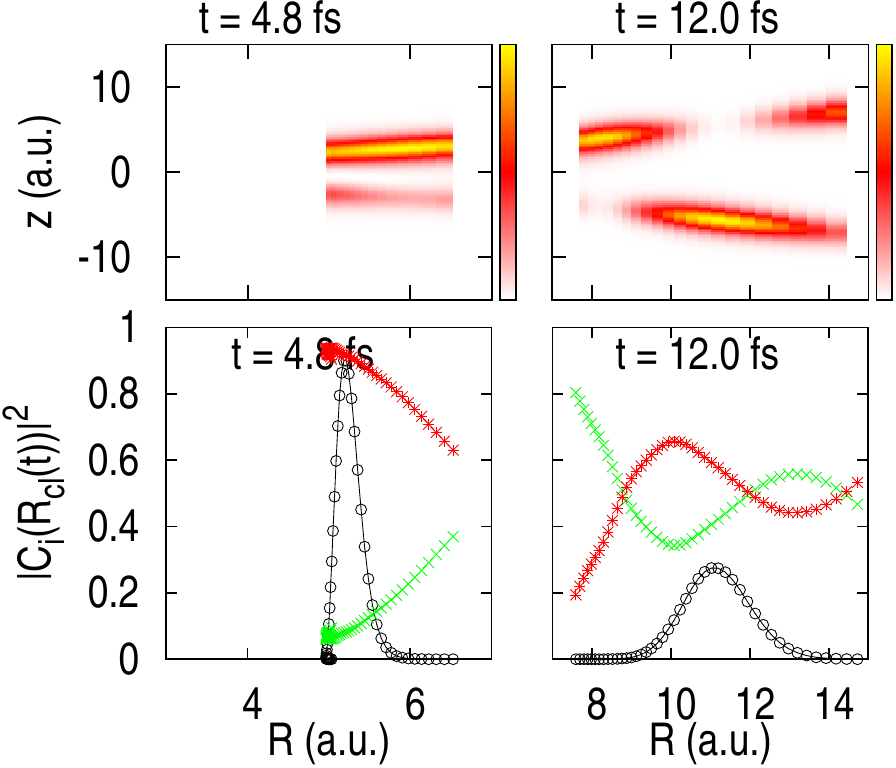}
 \caption{Upper panel: Electron density $|\Phi(z,t|R_{cl}(t))|^2$
 obtained from multiple trajectory Ehrenfest dynamics calculation at the indicated times (plotted for all trajectories $R_{cl}(t))$).
 Lower panel: Squared expansion coefficients of the Born-Oppenheimer expansion
 $|C_g(R_{cl}(t))|^2$ (green) and $|C_u(R_{cl}(t))|^2$ (red) of the electronic wave 
 function $\Phi(z,t|R_{cl}(t))$ obtained from multiple trajectory Ehrenfest dynamics calculation
 ($\Phi(z,t|R_{cl}(t))=C_g(R_{cl}(t))\Phi^g_R(z)+C_u(R_{cl}(t))\Phi^u_R(z)$) at the indiacted times. 
Nuclear density reconstructed from the distribution of 
classical trajectories are also plotted (black circle line). }
 \label{fig:Fig6}
\end{figure}

Given that there are several avoided crossings during the localization
dynamics, one might ask how well a mean-field surface to propagate the
electrons would work. To this end, we run a multiple-trajectory
Ehrenfest calculation~\footnote{A set of 800 trajectories is 
propagated according to 
\ben
\mu_n\frac{d}{dt}v_{cl}(t)=-\int dz\Phi(z,t|R_{cl}(t))(\frac{d}{dR}\hat{H}^{\rm int}_R)\Phi(z,t|R_{cl}(t))
\een
and
\ben
i\frac{\partial}{\partial t}\Phi(z,t|R_{cl}(t))=\hat{H}^{\rm int}_R(z,t)\Phi(z,t|R_{cl}(t)),
\een
where the initial conditions are sampled from the phase-space distribution
corresponding to $|\chi(R,t=0)|^2$.
}, and compare the electron and
nuclear densities with the exact ones.

In the upper panel of Fig.~\ref{fig:Fig5}, we plot
the conditional electron density $|\Phi_R(z,t)|^2$ obtained from the
exact calculation at indicated times.  In the lower panel, we 
plot its squared expansion coefficients of the Born-Oppenheimer
expansion $|C_g(R,t)|^2$ (green) and $|C_u(R,t)|^2$ (red) 
($\Phi_R(z,t)=C_g(R,t)\Phi^g_R(z)+C_u(R,t)\Phi^u_R(z)$), along with the nuclear density (black). 
In Fig.~\ref{fig:Fig6}, we plotted electron density $|\Phi(z,t|R_{cl}(t))|^2$ obtained from the
multiple trajectory Ehrenfest dynamics calculation at the indicated
times (plotted for all $800$ trajectories $R_{cl}(t))$.  The lower panel shows
the squared expansion coefficients of the
Born-Oppenheimer expansion $|C_g(R_{cl}(t))|^2$ (green) and
$|C_u(R_{cl}(t))|^2$ (red) of the electronic wave function
$\Phi(z,t|R_{cl}(t))$ obtained from multiple trajectory Ehrenfest
dynamics calculation
($\Phi(z,t|R_{cl}(t))=C_g(R_{cl}(t))\Phi^g_R(z)+C_u(R_{cl}(t))\Phi^u_R(z)$).
We also show the nucler densities reconstructed from the distribution of 
classical trajectories obtained from multiple trajectory Ehrenfest
dynamics calculation (black circle line).

Comparing the top panels of these figures shows that Ehrenfest
dynamics gets the overall structure of the electronic conditional
probability reasonably well, however not exactly; in fact, these
differences lead to an incorrect prediction of the localization
asymmetry. For example, at $t=12.0$ fs at the internuclear separation
where the nuclear density is peaked, the projections shown in the
lower panels, onto the BO surfaces predicted by Ehrenfest are each
close to 0.5, while the exact are closer to 0.6 and 0.4.  Given the
nature of the BO $g$ and $u$ states in terms of the left and right
basis (Sec.~\ref{sec:QSPES}), this suggests the localization asymmetry
predicted by Ehrenfest is close to 1:0 while the exact is close to
0.8:0.2. Indeed this is verified by the calculation of the asymmetry.
Further, throughout the width of the nuclear wavepacket, the Ehrenfest
projections remain close to 0.5, while the exact projections fall
away, indicating there is a larger degree of decoherence in the exact
dynamics, missed in the Ehrenfest dynamics.  The differences in the
conditional wavefunction and the BO projections is even greater where the
nuclear density is small ($R=8\sim10$ and $R=12\sim14$).

In the field-free problem of non-adiabatic charge-transfer~\cite{AAYG, AAYG2, AAYMMG, Min}, multiple-trajectory Ehrenfest
dynamics failed, and this might have been expected given that
the density spatially separates (branches) onto two different BO
surfaces. In the present case, the nuclear density does not
split in space, and actually predicts the nuclear dynamics quite well,
but the errors in the electronic dynamics are more
significant. Further, it is the same potential that evolves the
electrons in the Ehrenfest calculation as in surface-hopping
calculations, and this same potential was shown to lack significant
structures that the exact potential acting on the electron subsystem ({\it e}-TDPES)
in Ref.~\cite{SAMYG} has.

\section{Conclusions and Outlook}
The TDPES and vector potential arising from the exact factorization of
the molecular wavefunction exactly accounts for the coupling to the
electronic subsystem as well as coupling to external fields and so it
is important to understand their structure, and to relate this to the
QSPES which is traditionally used, in order to be able to develop
accurate practical mixed quantum-classical methods for strong-field
dynamics.  In this paper, we have studied the topical phenomenon of
laser-induced electron localization in the dissociation of H$_2^+$,
choosing a gauge where the TDPES is only potential acting on the
nuclear system.  We found that the gauge-independent component of the
TDPES has a mean-field-like character very close to the
density-weighted average of the QSPESs and yields an oscillatory force
on the nuclei.  The gauge-dependent component of the TDPES smoothens
the oscillations of the gauge-independent component and together they
lead to the correct force.

We demonstrated that running an ensemble of classical nuclear
trajectories on this exact TDPES accurately reproduces the exact
nuclear dynamics.  We found that the force obtained by considering
surface-hopping transitions between QSPESs at the laser-induced
avoided crossing approximates this exact force, after the localization
begins to set in.  We showed that errors in multiple-trajectory
Ehrenfest dynamics are less significant for the nuclear dynamics than
for the electronic dynamics explored in Ref.~\cite{SAMYG}, where it
was shown that Ehrenfest yields an incorrect electron localization
asymmetry. It is worth noting that the potential acting on the
electrons in Ehrenfest dynamics and in surface-hopping schemes lack
important step and peak features that the exact potential acting on
the electronic system (the {\it e}-TDPES) has. Therefore the results
of this study show that to reproduce the laser-induced electron
localization dynamics accurately by means of a mixed quantum-classical
dynamics scheme, we have to go beyond the traditional methods such as
surface-hopping or Ehrenfest methods.  Our results here encourage the
development of mixed quantum-classical schemes based on Eqs~(\ref{eqn:
  exact electronic eqn}) and (\ref{eqn: exact nuclear eqn})~\cite{Min}
to simulate strong-field processes.

{\it Acknowledgments:}
Partial support from the Deutsche Forschungsgemeinschaft (SFB 762),
the European Commission (FP7-NMP-CRONOS), and the 
U.S. Department of Energy, Office of Basic Energy Sciences, Division of Chemical Sciences, 
Geosciences and Biosciences under award DE-SC0008623 (NTM),is gratefully acknowledged. 

\bibliography{./localization}

\end{document}